# The End Restraint Method for Mechanically Perturbing Nucleic Acids *in silico*

Jack W Shepherd[a,b]* and Mark C Leake[a,b]

[a] Department of Physics, University of York, York, YO10 5DD

[b] Department of Biology, University of York, York, YO10 5DD

* To whom correspondence should be addressed. E-mail jack.shepherd@york.ac.uk

**Abstract**

Far from being a passive information store, the genome is a mechanically dynamic and diverse system in which torsion and tension fluctuate and combine to determine structure and help regulate gene expression. Much of this mechanical perturbation is due to molecular machines such as topoisomerases which must stretch and twist DNA as part of various functions including DNA repair and replication. While the broad-scale mechanical response of nucleic acids to tension and torsion is well characterized, detail at the single base pair level is beyond the limits of even super-resolution imaging. Here, we present a straightforward, flexible, and extensible umbrella-sampling protocol to twist and stretch nucleic acids *in silico* using the popular biomolecular simulation package Amber – though the principles we describe are applicable also to other packages such as GROMACS. We discuss how to set up the simulation system, decide forcefields and solvation models, and equilibrate. We then introduce the torsionally-constrained stretching protocol, and finally we present some analysis techniques we have used to characterize structural motif formation. Rather than define forces or fictional pseudoatoms, we instead define a fixed translation of specified atoms between each umbrella sampling step, which allows comparison with experiment without needing to estimate applied forces by simply using the fractional end-to-end displacement as a comparison metric. We hope that this easy to implement solution will be valuable for interrogating optical and magnetic tweezers data on nucleic acids at base pair resolution.

**1 Introduction**

*1.1 DNA mechanics* in vivo

In the living cell, DNA is a dynamic and constantly modulated system which interacts with itself as well as with other agents such as molecular machines and transcription factors. Regulating this are two primary mechanisms – sequence recognition and changes in structural conformation. In the former case, specific nucleotide sequences are recognised by transcription proteins as the start of a gene which is then read until the so-called "stop codon" is reached. However, whether this is possible is regulated by the overall configuration of the genome; genes can be switched on and off either by physicochemical means such as methylation or by burying the gene in the middle of the nucleus such that it cannot be read. Indeed, these dual regulation mechanisms may explain the long-standing mystery of why multiple copies of the same gene exist in multiple parts of the genome. Conversely, it has been shown that under mechanical stress, some transcription factor binding sites form a stable conformation that may be used to promote transcription factor recognition even in mechanically perturbed conditions. It is therefore vital to understand the effect of torsion and tension on the structure of DNA, be it in the live cell [1], *in vitro* or computationally [2], to elucidate genome expression. Moreover, as synthetic DNA and protein-based molecular machines are developed, knowing how DNA functions as a *material* rather than as a biological agent will become of some importance.

*1.2 Experimental methods*

DNA's mechanical properties have been of interest ever since its structure was definitively determined by Rosalind Franklin and colleagues. Initially, these took the form of fibre diffraction experiments in which purified DNA was semi-dehydrated, and imaged using X-ray diffraction, which led to several structures of DNA and DNA-ligand systems being elucidated [3]. However, the constraints imposed by the sample preparation and the static nature of the experimental technique meant that possible experiments were limited to static structure determination rather than time-resolved imaging of ongoing biomolecular processes.

In the 1980s however, two types of "tweezers" were developed using single-molecule biophysics techniques [4], which revolutionized the study of DNA as a material and led to greater insights in the physics of life at the single molecule level [5]. The first type, known as magnetic tweezers, consists of a permanent or electro- magnet on the outside of a sample. DNA is then tethered to a magnetic microbead at one end and at the other to the sample surface, or to another bead which has been immobilized on the cover slip. The magnet can then be moved, or the magnetic field rotated using electromagnetic coils [6], to exert a force on the bead, or twisted to exert torque. While this goes on, the DNA may be imaged using intercalating fluorescent dyes [7], or the position of the bead (and hence absolute force applied) can be tracked using brightfield imaging and centroid detection.

The second tweezers is known as the optical or laser tweezers and uses typically a latex or polystyrene bead which is trapped using a near-infrared laser. The laser is focused by the objective lens, and at the focus the radiation pressure of the light diffracting through the bead is sufficient to exert a Hookean force towards the center of the focus. Again, DNA can be tethered to the bead and at the other end to the cover slip or to a second bead which may also be trapped in a so-called "dumbbell" assay. The DNA may be imaged with fluorescence microscopy, or its structure inferred from force-extension behaviour, or both.

These tweezers have led to many important discoveries including the force-extension behaviour of DNA [8], of supercoiled DNA [9], and of DNA/protein systems [10]. DNA has been hyper-stretched [11], and overwound to form Pauling-like conformations [12]. When using fluorescence microscopy, colour-changing dyes may be used to measure the proportion of DNA which is single-stranded (melted) or remaining in the familiar double-helix structure [13], [14]. Fluorescence imaging can enable insight into DNA architecture and dynamics by imaging proteins that bind to DNA using live cell microscopy, such as in studying DNA replication [15], [16], remodelling [17], and DNA topology manipulation [18], but can be used *in vitro* in multi-spectral imaging, for example to DNA repair proteins translocating along DNA that has been tagged using a different colour dye [19], or it can be combined with other quantitative techniques such as fluorescence polarization which can infer the orientations of the base pairs in which the dye is intercalated [20]–[22]. These approaches, along with high-precision structural imaging methods such as atomic force microscopy (AFM) [23], have demonstrated that DNA is a complex and dynamic system where multiple competing motifs can coexist depending on the stress and torsion applied, and through inference the likely structures have been determined.

While these approaches are extremely powerful and have elucidated some of DNA's mechanical properties, there are some levels of detail which are unreachable through optical microscopy means. Thanks to the diffraction limit, individual base pairs cannot be imaged directly. Instead, we turn to fluorescent reporter molecules which are excited by a given wavelength of laser light. We can then localize these molecules individually as long as they are sufficiently separated in time or space. The localization precision possible though this method is generally of the order 20-30 nm [7], well below the diffraction limit but still representing almost 100 base pairs, while the dye itself is generally bound between 4-6 base pairs [24]. There is therefore a clear mismatch between the length scale that the molecule is actually reporting on and the length scale it is taken to be reporting on. Moreover, binding a dye in either groove or intercalating it between base pairs has significant structural implications; in the case of a commonly used intercalating dye YOYO- 1 for example the contour length can increase by up to 50% with associated underwinding and torsional stiffening of the double helix [24], [25], while it is unknown the specific mechanical stabilization or destabilization effects of an intercalator. It is therefore something of an open question as to how close to the unlabelled structure a fluorescently-labelled tract of DNA will be.

*1.3 Simulation approaches*

Fortunately, as computers have grown more powerful and general-purpose graphics processing units (GPGPUs) have developed, larger time- and length-scale simulations have become possible at both an atomistic level (using for example Amber [26]) and a coarse-grained level (for example using oxDNA [27]). Various strategies have been devised to mechanically perturb DNA systems using each regime which we summarize here.

First, it is possible to define a so-called "repulsion plane" in a simulation cell. This is a fictional barrier which repels a given subset of simulated atoms using a desired potential, generally either linear or harmonic. This can be used to stretch nucleic acids by placing the plane in the center of the contour and repelling the two ends. Alternatively, fictional "pseudoatoms" can be defined. These do not participate in the overall simulation but are placed away from the nucleic acid fragment towards the edge of the simulation cell and are used as anchors, held fixed in place while specified atoms are harmonically attracted towards them with a potential well. Both of these approaches have been successfully used to show DNA melting behaviour [28], but neither is well suited to torsional constraint.

The repulsion plane can be modified however using an additional co-ordinate, usually defined as the twist angle between the first and last base pairs [29]. An additional forcefield parameter can then be specified which restrains this angle to be in a specified range, essentially exerting torsional control, though this has the disadvantage of being difficult to implement. Other simulations have been run using an infinite length of DNA – that is, a fragment which joins itself at the edge of the periodic simulation box. Here, the DNA is torsionally constrained thanks to it being bound to itself, and base pairs were deleted to retain the same supercoiling density in fewer base pairs, in essence increasing the overtwisting [30]. This produced Pauling-like DNA (P-DNA) in which the base pairs are exposed and the backbone tightly wound around itself with a helical pitch of just a few base pairs. However, the utility of this approach beyond simulations of those types is limited.

Also using the DNA-bound-to-itself method of torsional constraint are minicircle based simulations [31]. These are circular DNA constructs the supercoiling density of which can be modified prior to simulation. Using these, Pyne, Noy and colleagues demonstrated the rich landscape of supercoiled minicircle DNA using both atomic-level simulations and AFM [32], demonstrating the power of combined theory and experiment studies. Minicircles, however, are not suitable for a single strand stretching experiment.

*1.4 Advantages of our approach*

Here we present our protocol for stretching DNA at a fixed supercoiling density by making use of atomic restraints on both terminal base pairs of a nucleic acid fragment and discrete movements of one end of the fragment to produce a stretch [33] (see a flow-chart representation of this in Figure 1). The advantages of this, we believe, are as follows: first, it is an exceptionally conceptually and practically straightforward method which is readily implemented in any simulation suite; second, because the end-to-end length of the nucleic acid is fixed, comparison with optical or magnetic tweezer assays is rendered more straightforward than force-matching approaches; third, because the method makes use of umbrella sampling, it is possible to parallelize across sampling windows to reduce overall runtime (see Figure 1); fourth it is trivial to perform these simulations with any desired supercoiling density, up to any desired overstretch; fifth, it is possible to run any desired energy minimization routine at the start of each umbrella sampling window, so for especially complex systems a full simulated annealing stage could be implemented if needed; sixth, by lowering the strength of the restraint potential on the terminal base pairs some movement of the ends is allowed – the end-to-end length can then be used as the co-ordinate for the weighted histogram method of free energy calculation (WHAM) and the change in free energy due to conformation change can be extracted; and seventh, this approach is trivially compatible with ligand binding.

To our knowledge, our protocol is unique in this list of advantages. We note however that there are some disadvantages of our approach. Most notably, each perturbation is a discrete single step, after which the structure is energetically minimized. We do not therefore access the transition between sub-conformations, we simply see them in their formed states. Because we artificially keep the terminal base pairs in a canonical conformation, care must be taken to use a sufficiently long fragment so that edge effects are not a concern at the centre, where analysis will then be focused. We also do not generally access the forces applied due to these restraints (it is possible to extract these values but in our experience they are invariably too noisy to be useful) so comparison with experiment at that level is largely impossible. However, for linear stretching simulations with torsional constraint these drawbacks are surmountable.

## 2    Materials

### 2.1 Molecular dynamics software

Here we focus on atomistic molecular dynamics using Amber and its associated forcefields. For these structurally perturbed simulations it is necessary to choose a modern well-tested forcefield and we in general use either the *OL15* or *bsc1* nucleic acid forcefield [34]. For solvation we use the Generalized Born model when working with implicit solvation and a TIP3P [35] model for explicit solvation, with appropriate ion parameters. Systems are prepared with the Amber utilities *nab* and *leap*.

### 2.2 Visualization software

For most atomistic operations we recommend Visual Molecular Dynamics (VMD) [36], a freely available biomolecule visualization suite which can be run using input scripts or through the visual interface. However, for presenting coarse-grained structures (eg with a whole nucleotide represented by a single flat cuboid) two alternatives are Ovito [37] and Chimera [38], both free and scriptable as well as operating in GUI mode.

### 2.3 Analysis utilities

The free AmberTools suite is usually sufficient for the analysis that is needed for these linear stretching simulations as writhe and plectoneme formation is prohibited by the simulation setup. Using AmberTools' *cpptraj* [39] we can find canonical and non-canonical hydrogen bonds, while with *sander* we can evaluate the total interaction energy between nucleotides. More difficult is conformational analysis as the resulting highly mechanically perturbed structures lead to anomalous results (for example, if a base pair melts the assumed vector linking the nucleotides ceases to be meaningful). However, for unmelted structures it is possible to use Curves+ [40]. Other physical properties such as the persistence length can be estimated using the *SerraNA* [41] utility which uses the molecular flexibility and correlations between fluctuations in different base pairs to estimate mechanical parameters. However, we note that the underlying assumption in *SerraNA* is that the ends of the DNA are free to fluctuate so the output for an end-constrained simulation will be only a rough estimate.

*2.4 Bespoke scripts and input files*

Various scripts will need to be written to perform the functions described below. Our own pipeline is composed of *bash, awk,* and Python 3 scripts chained together in one *bash* script file which also calls the necessary *tleap* and *cpptraj* instances during the simulation. In general we recommend developing these scripts in pure Python which will be most portable and maintainable. Input files to Amber will need to be written on a case-by-case basis though we give some examples below.

## 3 Methods

*3.1 Initial system preparation*

1. Create a linear piece of DNA using the *fd_helix* and *putpdb* functions in *nab*. This by default lays the helical axis along the simulation *z* axis which will be useful for implicit solvent simulations. For ease we will call the resulting pdb structure **nab_structure.pdb**
2. If applying no additional supercoiling, prepare input files for Amber's *pmemd.cuda.* To do this, it is best to write a reusable *tleap* script. In the case of an implicit solvent simulation this can be as simple as the following:

    ```
    a = loadpdb("nab_structure.pdb")
    saveamberparm a nab_structure.prmtop nab_structure.inpcrd
    ```
    Save this as **tleap.in** and run as follows: $ tleap –f tleap.in This will create two files - **nab_structure.prmtop** and **nab_structure.inpcrd** which can be given to *pmemd.cuda* for simulation.

    If using TIP3P water, the input file is only slightly more complex. For an octahedral simulation cell with a size 12x12x50 Å:

    ```
    a = loadpdb("nab_structure.pdb")
    solvatebox a TIP3PBOX {12 12 50}
    addions a Na+ 0
    saveamberparm a nab_structure.prmtop nab_structure.inpcrd
    ```
    In order to avoid having to resolvate the structure at any point during the stretching simulation, it is imperative to select a box large enough that the stretched DNA will still fit inside it comfortably, with ~12 Å clearance between the box edge and any atom in the DNA. Here note also that we have included a line which adds Na+ ions so that the overall system is electrically neutral. This is essentially proceeding at 0 M salt concentration – hardly biologically relevant. See Notes for adding extra ions.

*3.2 Applying torsion*

Application of torsion is of course nothing more than rotating the terminal base pair around the helical axis to a desired supercoiling density. This is most easily achieved immediately after preparation of the **nab_structure.pdb** file when the helical axis lies along *z.* Read in the pdb file to your scripting language of choice and simply apply the 3D rotation matrix to rotate by some step around *z.* For large rotations, it will be necessary to perform energy minimization between smaller rotation events to build up a large supercoiling density.

Having done this, you will need to prepare the *pmemd.cuda* input files as described in 3.1 above.

*3.3 Equilibration*

Equilibration is reasonably straightforward and proceeds according to several well-known protocols. For the purposes of the input file examples below we will assume that the simulation duplex is 20 base pairs long, therefore containing 40 nucleotides in total.

1. If using explicit solvent, begin by performing steepest descent and conjugate gradient on the water and ions only, keeping the DNA fixed in place. Example input file:
   ```
   &ctrl
     imin=1, ncyc=1000, maxcyc=2000, ntb=1, igb=0, cut=12, ntr=1
   /
   DNA restraints
   50.0
   RES 1 40
   END
   END
   ```
2. Perform steepest descent and conjugate gradient minimization on the middle of the NA fragment, keeping end base pairs only fixed. We find that this step is crucial – in general we use up to hundreds of thousands of minimization steps to ensure that the structure is as close as possible to its ground state. Example input file:
   ```
   &ctrl
     imin=1, ncyc=10000, maxcyc=20000, ntb=1, igb=0, cut=12
   /
   DNA end restraints
   50.0
   RES 1,20,21,40
   END
   END
   ```
3. If using explicit solvent, perform a 100 ps simulation in the NPT ensemble, heating the system from 0 to 300 K linearly through the simulation. However, if using implicit solvent this step is unnecessary for simple simulations though would likely be useful for complex nucleic acid/ligand systems.

*3.4 Running the umbrella sampling simulations*

Finally the time has come to perform the umbrella sampling simulations themselves. We proceed according to the flowchart in Figure 1.

1. First, simulate the unstretched conformation making sure to keep the end restraints in place. This can be done with the user's preferred parameters, though we recommend keeping the timestep small (≤2 fs) and using the maximum velocity keyword (maxvel) to limit the maximum atomic velocity to *ca.* 20. Physically the need for this arises thanks to the hard restraints on the end base pairs – if a water atom gets too close to this hard wall, it will be repelled at a very high force which may lead to the simulation blowing up energetically. The need for this is lessened by using softer restraints on the end base pairs, though for consistency if nothing else it is worthwhile deciding on a maximum velocity and sticking with it. Ensure to use the positional restraints on the terminal base pairs.
2. While the molecular dynamics goes on, monitor the various physical parameters that indicate it has reached thermodynamic equilibrium – for example density and total energy.

At the beginning of the simulation the system will still be equilibrating so there will be a number of frames that need to be discarded or at least ignored for the purposes of analysis.
3. When thermodynamic equilibrium appears to have been reached, you may then take a snapshot from the system by copying the current restart (.rst) file to a new directory. This will be the starting conformation for the simulation of the first stretching event.
4. How to proceed here depends on whether you are using explicit or implicit solvation. In the table below we summarize how to approach each.

| Implicit solvation | Explicit solvation |
|---|---|
| - *Here, the easiest way to proceed is to use cpptraj to create a new pdb file, for example: $ cpptraj –p dna.prmtop –y dna.rst –x dna.pdb*<br>- *This pdb file will still have the helical axis along z*<br>- *It is therefore trivial to read in the pdb file, and subtract 1 from the z component of position for the terminal nucleotides by using the residue number*<br>- *This can be done in a variety of ways but the most straightforward may be using awk:*<br><br>```
{if($5 == 1 || $5 == 40){
   printf("ATOM %6i %4s %3s %5i %8.3f%8.3f%8.3f  1.00  0.00\n", $2, $3,\
 $4, $5, $6, $7, $8-1);
  else{
    print $0
  }
}
```<br>- *If saved as e.g. stretch.awk this script can then be run as follows:*<br>`$ awk –f stretch.awk < dna.pdb > dna_stretched.pdb`<br><br>- *This dna_stretched.pdb can then be used as the starting pdb for minimization and simulation as described above* | - *If explicit solvation is being used, two methods may be employed. Most straightforwardly, a similar .pdb file may be created using cpptraj as described for the implicit case*<br>- *awk may then also be used to move the terminal base pair by checking residue number*<br>- *However, in our experience the pdb files often fail to encode the box information properly*<br>- *It is therefore necessary to redefine the box using tleap when it is loaded in to create the topology and co-ordinate files using the* setbox *command*<br>- *Alternatively, the .rst file can be modified directly if it is saved as a plain text netCDF file*<br>- *The easiest approach to this is to read the file in using Python, then move co-ordinates for the correct residues, though the netCDF does not specify residue numbers – these can be taken from the topology file*<br>- *For explicit solvent simulations, especially if using a truncated octahedral box rather than a cuboid, it is not guaranteed for the DNA helical axis to lie along z however*<br>- *It is therefore in general necessary to specify a translation axis for the terminal base pair, which may be worked out from atomic co-ordinates of the nucleic acid.* |

5. Having stretched the nucleic acid by 1 Å, this is used as the starting structure for minimization and simulation as described above.

*3.5 Analysis strategies*
1. Finally, the overall simulation can be analyzed either in individual umbrella sampling windows or by joining the windows into one simulation – making sure to discard the start of each window which may not be in equilibrium.
2. Use Curves+ and SerraNA to estimate physical properties, though note than in particularly perturbed states these estimates will likely be unreliable.

3. In general there are two useful properties for observing motif formation: hydrogen bonding and stacking energy
4. To calculate the hydrogen bonding, we generally use cpptraj with a 120° angle cutoff and a 3.5 Å distance cut-off
5. This produces a large table of hydrogen bonding interactions, and using Python or another scripting language, the canonical and non-canonical hydrogen bonds can be extracted
6. For the stacking energy, the AmberTools utility *sander* is used. However, *sander*'s *esander* routine does not decompose energies by residue-residue interaction so the user must do this manually
7. The easiest way – though time consuming to do – is simply to iterate through each possible pair of nucleotides, strip out all others and run esander. This comes at some computational cost, however. For canonical base stacking between only nearest neighbours, it is relatively cheap – N operations where N is the number of nucleotides in the system. However, for noncanonical stacking, in principle this would take (N-1)(N-2)/2 operations, quickly becoming prohibitively expensive (for 40 nucleotides this is 741 separate calculations). However, we find that with a linear fragment of DNA being stretched we can in practice limit the operations to next nearest neighbours, including cross-strand neighbours- i.e. the three nearest neighbours on the opposite strand plus the two next-nearest neighbours for a total of 5 per nucleotide, though these are degenerate – each calculation will be used for two nucleotides. The number of calculations is therefore 5N/2, a significant saving.
8. Alternatively, it is possible to find the pairwise energies by using other postprocessing analysis software – GROMACS does have this functionality but it comes with a necessary file conversion step.
9. Either way, by calculating these energies, multiple files will be generated. From these, for each nucleotide, it is possible to extract the nonbonded energy (principally Van der Waals interactions) and sum all contributions for each nucleotide. The total canonical or noncanonical stacking is thus known.
10. If a high noncanonical stacking energy or number of hydrogen bonds is seen to persist over many frames it is indicative of a semi-stable structure being formed. The easiest way to see the trends in these physical quantities is by plotting a heatmap in which the *x* axis is base pair, the *y* axis is the time axis, and the color corresponds to the physical quantity under examination.
11. All apparent structures should then be interrogated with visualization software and the precise details obtained from the hydrogen bond and energy analysis.
12. Finally, if soft restraints are used on the terminal base pairs, the free energy may be estimated using WHAM. In order to do this, the integration coordinate must first be determined from the trajectory files. In general we have had good success using the end-to-end length of the DNA as the coordinate, found trivially from the atomic *xyz* positions.

## 4 Notes

### 4.1 Ionic strength

1. If using implicit solvation, set the salt concentration using the saltcon keyword in the *pmemd.cuda* input file. This is measured in moles/L so saltcon=0.2 would generate Debye-Huckel screening parameters to approximate 200 mM salt.
2. If using explicit solvation, the user must solvate the system and note the volume of the bounding box. From this one can work out the required number of salt ions needed and add them manually after adding enough ions to electrically neutralize the system.

3. Imagining that the number of ions needed was 20, we could proceed as follows:
   ```
   a = loadpdb("dna.pdb")
   solvatebox a TIP3PBOX {15 15 50}
   addions a Na+ 0
   addions a Na+ 20
   addions a Cl- 20
   saveamberparm a dna.prmtop dna.inpcrd
   ```
4. Of course this process is also scriptable, and includes calling *tleap* once to generate the solvated system, then saving as a pdb file, extracting the volume from the automatically generated leap.log logfile, and then loading the solvated system back in to *tleap* to add the ions before saving the topology and coordinate files for simulation.

**Acknowledgements**

Thank you for helpful discussions with Agnes Noy, Matt Probert, and Robert Greenall at the University of York. Work was supported by the Leverhulme Trust (RPG-2019-156, RPG-2017-340) and EPSRC (EP/N027639/1).

**Figures and captions**

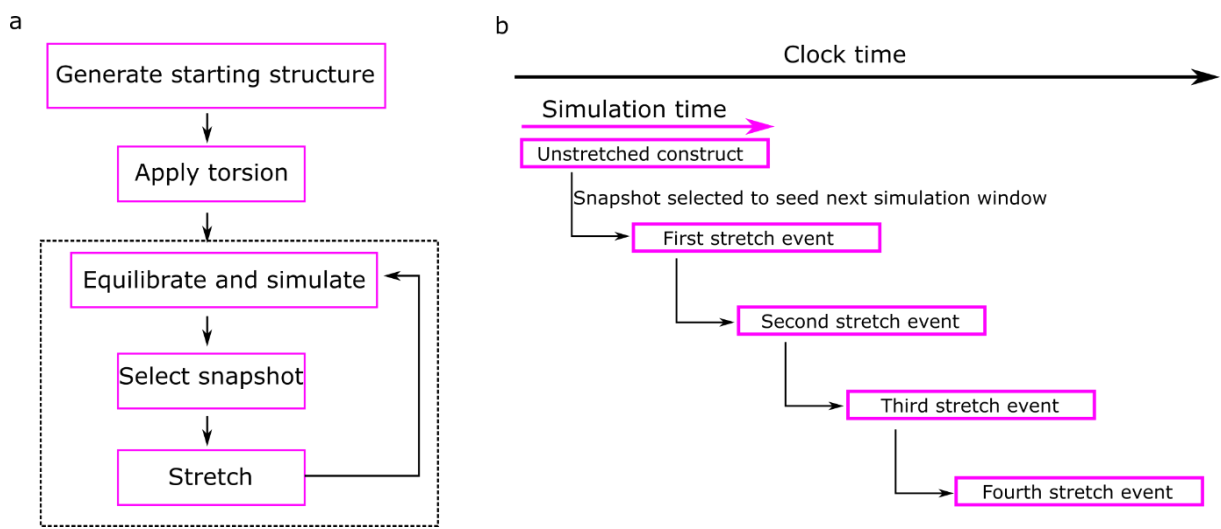

**Figure 1:** a) Flowchart of the end-restraints method for nucleic acid perturbation. The steps inside the dashed box must be repeated to build up a large overall stretch in many small increments. b) Each stretch even seeds the next, and by seeding step *n+1* before step *n* is finished, the simulation can be efficiently parallelized across umbrella sampling windows.